\documentclass{article}
\usepackage{spconf}
\usepackage{graphicx}
\usepackage{subcaption}
\usepackage{hyperref}
\usepackage{amsmath}
\usepackage{multirow}

\usepackage[font={small}]{caption}
\usepackage[ruled,linesnumbered]{algorithm2e}
\SetKwRepeat{Do}{do}{while}

\title{Spectral Data Augmentation Techniques to quantify Lung Pathology from CT-images}
\name{Subhradeep Kayal$^{1}$, Florian Dubost$^{1}$, Harm A. W. M. Tiddens$^{2}$, Marleen de Bruijne$^{1, 3}$}
\address{$^{1}$ Biomedical Imaging Group Rotterdam, Erasmus MC Rotterdam, The Netherlands \\
     $^{2}$ Department of Pediatric Pulmonology and Allergology, Erasmus MC Rotterdam, the Netherlands \\
     $^{3}$ Department of Computer Science, University of Copenhagen, Denmark}
\begin{document}
%
\maketitle
\begin{abstract}
Data augmentation is of paramount importance in biomedical image processing tasks, characterized by inadequate amounts of labelled data, to best use all of the data that is present. In-use techniques range from intensity transformations and elastic deformations, to linearly combining existing data points to make new ones. In this work, we propose the use of spectral techniques for data augmentation, using the discrete cosine and wavelet transforms. We empirically evaluate our approaches on a CT texture analysis task to detect abnormal lung-tissue in patients with cystic fibrosis. Empirical experiments show that the proposed spectral methods perform favourably as compared to the existing methods. When used in combination with existing methods, our proposed approach can increase the relative minor class segmentation performance by 44.1\% over a simple replication baseline.
\end{abstract}

\section{Introduction and Motivation}

Learning tasks in the biomedical domain are often constrained by the lack of substantial labelled data and the presence of imbalance problems: be it class imbalance in healthy-vs-diseased image classification, or imbalance in the number of pixels corresponding to healthy and diseased areas in semantic segmentation. In such cases, \emph{data augmentation techniques}, which can be used to artificially increase the size and balance the training dataset, can help regularize a learning model thereby increasing the generalization performance.

Several effective methods for data augmentation have been proposed in recent years, ranging from simple methods such as flips, rotations, scaling, local zooming \cite{GIBSON2018113}, additive Gaussian noise \cite{simpleaug} and intensity transformations \cite{intensity}, to elastic deformations, which can mimic the possible movements of biological tissues \cite{unet}. Zhang et al. \cite{mixup} proposed to make new image datapoints by randomly selecting two existing images and linearly combining them, which was extended by Eaton-Rosen et al. \cite{mixmatch} to include the class distribution information when picking two images to combine. Generative adversarial networks have also been used for data augmentation during network training in \cite{rendergan}, but are expensive. A useful review comparing various data augmentation has been provided by Perez et al. in \cite{summary}.

Most of the aforementioned methods work directly on the spatial domain, i.e. on the pixel values directly. In this paper, synthetic samples for data augmentation are created based on the frequency (or spectral) domain information, thus being complementary to other methods in existing literature. In particular, we exploit the discrete cosine and wavelet transforms as they achieve high-fidelity in signal reconstruction while being computationally efficient. The said transforms are used to decompose an image into its bases, and recompose the bases into synthetic images, by adding some corrupting noise to them. This results in a set of images with slightly altered spectral components and therefore slightly different noise properties, mimicking different CT scan protocols. We evaluate the proposed approaches of data augmentation in a U-net \cite{unet} based system to quantify lung pathology in CT scans of patients with cystic fibrosis.

\section{Methods} 

\subsection{Spectral Methods for Data Augmentation}

In this work, we exploit two particular methods to transform an image into its spectral components. The first, \emph{Discrete Cosine Transform (DCT)} \cite{dct}, aims at decomposing an image into a sum of infinite-range cosine basis functions oscillating at different frequencies, which are the DCT components of the parent image.

For a 2D \begin{math} N \times M \end{math} image \begin{math} f \end{math}, the 2D DCT is given by
\begin{equation}
\small
\begin{split}
F(U,V) = (\frac{2}{N})^{\frac{1}{2}} (\frac{2}{M})^{\frac{1}{2}} \sum_{i=0}^{N-1}\sum_{j=0}^{M-1}\nu(U)\nu(V) \\
\cos(\frac{\pi U}{2N}(2i + 1))\cos(\frac{\pi V}{2M}(2j + 1)).f(i, j)
\end{split}
\label{DCT}
\end{equation}
where \begin{math} F \end{math} is the 2D DCT components matrix and, \begin{math}\small \nu(\epsilon)=
\begin{cases}
    \frac{1}{\sqrt{2}},& \text{if } \epsilon = 0\\
    1,              & \text{otherwise}
\end{cases}
\end{math}
The inverse transform has the same form as Equation \ref{DCT}, with \begin{math} F(U,V) \end{math} and \begin{math} f(i,j) \end{math} exchanging places in the equation. For simplicity, we write it as,
\begin{equation}
\small
\label{idct}
f(i, j) = F^{-1}(U,V)
\end{equation}

We apply an additive\footnote{A multiplicative Gaussian noise was also experimented with, but did not produce reasonable results.} corrupting Gaussian noise to the DCT components matrix, produced by equation \ref{DCT}, such that the noise added is in proportion to the value of component \begin{math} F(U,V) \end{math}. Thus, lower frequencies, which store the bulk structural information of the image, are less susceptible to modification than higher frequencies, storing the more fine-grained information. This is represented in the following equation:
\begin{equation}
\small
\label{dctsynth}
F'(U,V) = F(U,V) +\mathcal{N}(0, \rho F(U,V))
\end{equation}
where, \begin{math} \rho \end{math} takes \begin{math} R \end{math} values linearly between 0 and \begin{math} \eta \end{math}, such that \begin{math} R \end{math} is the desired number of replications and \begin{math} \eta \end{math} is a fraction controlling the maximum noise added. For example, if \begin{math} R=3 \end{math} and \begin{math} \eta = 0.3 \end{math}, meaning that we want to produce 3 synthetic images, with the maximum corrupting noise proportional to 30\% of the strength of a component, then \begin{math} \rho \end{math} takes values of 0.1, 0.2 and 0.3. Finally, we apply equation \ref{idct} to the produced components to retrieve the synthesized images.

The second transformation that is utilized in this paper is the \emph{Discrete Wavelet Transform (DWT)} \cite{dwt}, which captures both the frequency and location information of an image being decomposed. This is because while DCT uses infinite-range cosine bases, DWT utilizes fixed-length custom basis functions for the decomposition of an image.

2D-DWT makes use of a scaling function, \begin{math} \phi \end{math}, and three wavelets, \begin{math} \psi^H \end{math}, \begin{math} \psi^V \end{math} and \begin{math} \psi^D \end{math}, where H, V and D signify wavelets which are directionally sensitive in the horizontal, vertical and diagonal directions. For scale \begin{math} s \end{math}, the scaled and translated basis functions are given as
\begin{align}
\small
\label{translatedscaled}
\begin{split}
 \phi_{s,U,V} (i,j) = 2^{\frac{s}{2}} \phi(2^s i - U, 2^s j - V), \\
 \psi_{s,U,V}^k (i,j) = 2^{\frac{s}{2}} \psi^k(2^s i - U, 2^s j - V)
\end{split}
\end{align}
where \begin{math} k = {H, V, D} \end{math}.

Considering again a 2D \begin{math} N \times M \end{math} image \begin{math} f \end{math}, and using equation \ref{translatedscaled}, the DWT components are defined as follows,
\begin{align}
\small
\begin{split}
 W_{\phi} (s_0, U, V) = \frac{1}{\sqrt{MN}}\sum_{i=0}^{N-1}\sum_{j=0}^{M-1} f(i,j) \phi_{s_0,U,V} (i,j)
\\
 W^k_{\psi} (s, U, V) = \frac{1}{\sqrt{MN}}\sum_{i=0}^{N-1}\sum_{j=0}^{M-1} f(i,j) \psi_{s,U,V}^k (i,j)
\end{split}
\label{DWT}
\end{align}
where \begin{math} k = {H, V, D} \end{math} and \begin{math} s_0 \end{math} is the starting scale, beginning with the raw image.
The component matrix \begin{math} W_{\phi} \end{math} contains the \emph{approximation coefficients}, whereas the \begin{math} W^k_{\psi} \end{math} have the \emph{detail coefficients}.

\begin{figure}[!tbp]
\setlength\belowcaptionskip{-5pt}
  \centering
    \includegraphics[width=0.48\textwidth]{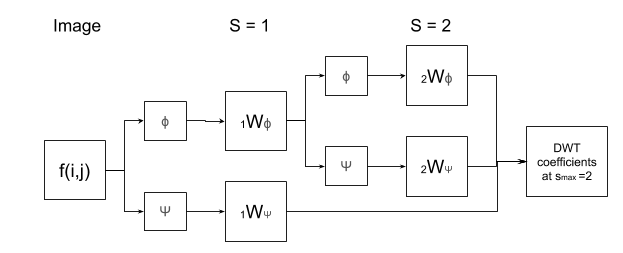}
  \caption{A schematic depicting a two-level DWT. $f(i,j)$ is the original image, $\phi$ and $\psi$ denote the scaling and wavelet functions respectively.}
  \label{dwtpic}
\end{figure}

The inverse DWT is then
\begin{align}
\small
\label{idwt}
\begin{split}
f(i,j) = \frac{1}{\sqrt{MN}}\sum_{U}\sum_{V} W_{\phi} (s_0, U, V) \phi_{s_0,U,V} (i,j)
\\
 + \frac{1}{\sqrt{MN}}\sum_{k = H,V,D}\sum_{s = s_0}^{s_{max}}\sum_{U}\sum_{V} W^k_{\psi} (s, U, V) \psi^k_{s,U,V} (i,j)
\end{split}
\end{align}
where \begin{math} s_{max} \end{math} is the maximum scale that is chosen.
A schematic of the DWT process is show in Figure \ref{dwtpic}.

In a manner similar to the use of DCT components in order to create synthetic images, we corrupt the detail and approximation coefficients in proportion to their magnitude.
\begin{equation}
\small
\label{dwtsynth}
\begin{split}
{W'}_{\phi} (s_0, U, V) = {W}_{\phi} (s_0, U, V) +\mathcal{N}(0, \rho {W}_{\phi} (s_0, U, V)) \\
{W'}^k_{\psi} (s, U, V) = {W}^k_{\psi} (s, U, V) +\mathcal{N}(0, \rho {W}^k_{\psi} (s, U, V))
\end{split}
\end{equation}
with the same definition of \begin{math} \rho \end{math} as aforementioned. Then the synthetic images can be recomposed by using equation \ref{idwt} on the corrupted coefficients as above.

A summary of steps can be found in Algorithm \ref{algo}.

\begin{algorithm}[t]
    \KwData{A 2-D image \begin{math} I \end{math}, number of desired replications \begin{math} R \end{math}, maximum corrupting noise \begin{math} \eta \end{math}}
    \KwResult{A set of \begin{math} R \end{math} images}

    Get either the 2D DCT components matrix \begin{math} F \end{math} (Equation \ref{DCT}), or the detail coefficients matrix \begin{math} {W}^k_{\psi} \end{math} (Equation \ref{DWT});

    Define \begin{math} \rho \end{math} as a set of \begin{math} R \end{math} linearly spaced values from 0 to maximum noise \begin{math} \eta \end{math};
    
    For every value of \begin{math} \rho \end{math}, corrupt \begin{math} F \end{math} or \begin{math} {W}_{\phi} \end{math}, \begin{math} {W}^k_{\psi} \end{math} using Equation \ref{dctsynth} or \ref{dwtsynth}, respectively;
    
    Add a synthetic image to the final set of images to return, by performing the inverse transform on the corrupted components, using either Equation \ref{idct} or \ref{idwt}, respectively;
    \caption{Constructing synthetic samples using \emph{DCT} or \emph{DWT}}
    \label{algo}
\end{algorithm}

\subsection{Convolutional Neural Networks for Lung Tissue Identification}
\label{unetdetails}

Most existing approaches for lung tissue pattern recognition are designed as a patchwise-classification scheme, ranging from the use of predefined filter banks as features \cite{ciompi}, to convolutional neural networks (CNN) based techniques \cite{marques}. Recently, semantic segmentation using CNNs have also been found to work well in order to identify pathological lung tissues \cite{anthimopoulos2}.
\begin{figure*}[!h]
\setlength\belowcaptionskip{-5pt}
  \centering
    \includegraphics[width=0.98\textwidth]{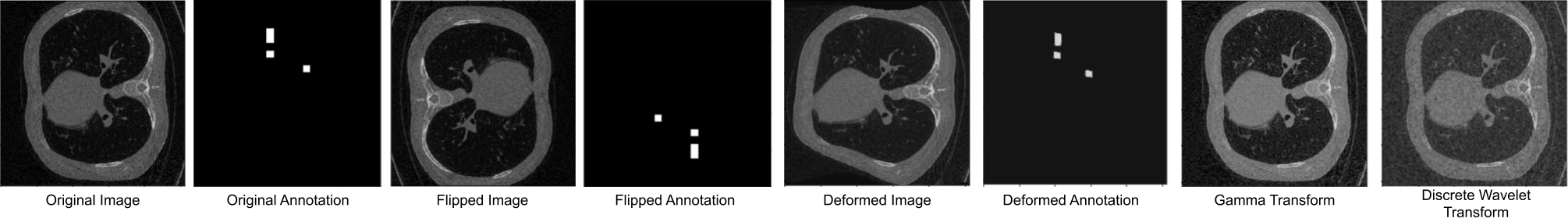}
  \caption{Original image, annotation and some examples of images synthesized via the various tested mechanisms. The gamma transformed image is created with $\gamma = 0.6$. The DWT synthesized image is with 30\% noise applied to the DWT components.}
  \label{augment}
\end{figure*}

In this work, we use a shallow 2D U-net architecture \cite{unet}, as shown in Figure \ref{cnnfig}. The major reason for using a 2D U-net architecture instead of a 3D one is the large variance in the number of slices from one CT scan to another in the dataset (See section \ref{data} for details). To optimize the network we minimize the \emph{Dice loss} using the \emph{Adagrad} \cite{adagrad} algorithm for optimization.

\section{Experiments} 

\subsection{Data}
\label{data}

The data used for this study was collected retrospectively, from the routinely acquired data of the Sophia Children’s cohort. This study was approved by the ethical review board of the Erasmus MC-Sophia Children’s Hospital (MEC2013‐338).

The dataset consists of annotated CT scans at full inspiration breath-hold, acquired from children aged between 1 year 11 months to 18 years old with early cystic fibrosis lung disease, and have slice-thickness ranging from 0.75 to 3 mm \cite{pragma}. After removing scans which have only 1 annotated slice, each scan was found to have between 9 and 11 patchwise annotated slices, with each patch being 20x20 pixels. For each patch that is at least 50\% covered by the lung field, the presence of an abnormality has been annotated according to the following system in order of priority: bronchiectasis, mucous plugging, bronchial wall thickening, atelectasis or normal lung structure. In our work, we do not discriminate between the individual abnormalities and label a patch containing any of the abnormal lung structure as a \emph{diseased} patch, whereas the absence of abnormality is marked as \emph{healthy}.

For this work, we split the dataset such that 131 CT scans form the training set, 33 are in the validation set and 42 images for the test set; the number of 2D images in each set are 1211, 329 and 412 respectively. The split is made such that the proportion of overall diseased and healthy patches remain constant across the sets, and no two patients are shared between the sets.

\subsection{Baseline Data Augmentation Methods}

As baselines we use:

\emph{simple} replication, where an image is copied as-is;

\emph{affine} transformations, which includes horizontal and vertical flips, rotation and scaling of an image patch;

\emph{intensity} based mechanisms, where we employ the gamma transformation with different values of gamma to create synthetic images;

\emph{elastic} deformations, where a coarse displacement grid is generated with a random displacement for each grid point and the input image is modified by using the displacement vectors and spline interpolation.

Some examples of augmented images are shown in Figure \ref{augment}.

\subsection{Experimental Settings}

\textbf{Environment:} The experiments are implemented in Python using the help of the Keras deep learning high-level library, and run on a Nvidia GeForce GTX 1070 GPU.

\begin{figure}[!h]
\setlength\belowcaptionskip{-5pt}
  \centering
    \includegraphics[width=0.48\textwidth]{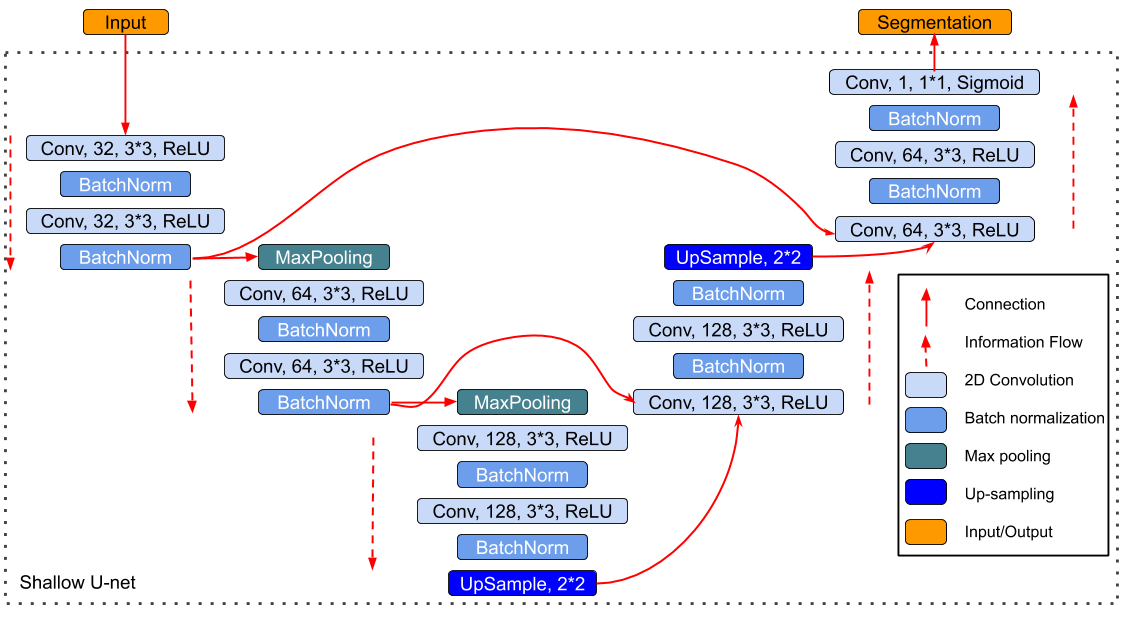}
  \caption{Architecture of the shallow U-net used. \emph{Conv, 32, 3*3, ReLU} indicates a convolution with 32 filters and filter-size 3x3 with \emph{ReLU} activation. \emph{UpSample, 2*2} indicates an up-sampling factor of 2.}
  \label{cnnfig}
\end{figure}

\textbf{U-net parameters:} For training the U-net, 512x512 input images are fed to it in batches of 3, which was the maximum batch-size to fit in the GPU memory. The learning rate is initiated at 0.01 and the number of epochs is set to 100, with the possibility of \emph{early stopping}, if the validation loss does not increase by at least 0.0001 over 20 iterations. Lastly, the reported results are an average of 5 runs for each experiment, such that the network is randomly initialized in each run. The internal parameters of the U-net are shown in Figure \ref{cnnfig}.

\textbf{Data augmentation parameters:} We study the effect of augmentation by replicating images containing diseased class pixels \begin{math} (R = )5 \end{math} times for all methods tested. In the context of the spectral transforms studied here, experiments were performed with maximum corrupting noise levels, \begin{math} \eta \end{math}, of 0.5\%, 1\%, 5\% and 10\%. For elastic deformations, a 4x4 grid was used to deform the image, with the displacement vector linearly varying from 1 to 20 pixels, while for affine transformations, each image can be horizontally or vertically flipped and rotated by 0 to 10 degrees randomly. Finally, for the intensity transforms, the value of gamma is varied linearly between 0.8 to 1.2.

\textbf{Post-processing:} Since the U-net provides a pixelwise segmentation map, while for this dataset we have patchwise labels, the output of the U-net is subjected to a morphological hole-filling step with a square mask of ones of size 5x5 pixels.

\textbf{Training set size:} Furthermore, in order to see the effectiveness of the augmentation methods in relation to the size of the training set, experiments were performed where training was done on 20, 60, 100 and finally, the whole training set of 131 CT scans, whereas testing is always performed on the 42 scans in the test set. The results are reported next.

\textbf{Evaluation:} We evaluate the proposed methods by calculating the \emph{F1-score} for the \emph{disease} class, since it is imperative that the diseased tissues are correctly identified in real situations.

\section{Results and Discussion}

\begin{table}[!tbp]
\centering
\begin{tabular}{c|cccc}
\textbf{Training Size}                                                                                & \textbf{20}                          & \textbf{60}                          & \textbf{100}                         & \textbf{All}                         \\ \hline
\textbf{Method}                                                                                       &                                      &                                      &                                      &                                      \\
simple                                                                                                & 0.12448                              & 0.14902                              & 0.21480                              & 0.26459                              \\
affine                                                                                    & 0.16539                     & 0.21057                     & 0.29174                     & 0.31673                     \\
intensity                                                                                             & 0.13604                              & 0.16334                              & 0.21881                              & 0.28683                              \\
elastic                                                                                      & 0.17392                     & 0.23458                     & 0.30989                     & 0.33656                     \\
                                                                                                      &                                      &                                      &                                      &                                      \\
DCT@0.5\%                                                                                             & 0.15794                              & 0.21013                              & 0.28390                              & 0.30622                              \\
DCT@\%                                                                                      & 0.17820                     & 0.23313                     & 0.30390                     & 0.33124                     \\
DCT@5\%                                                                                               & 0.14946                              & 0.20465                              & 0.27515                              & 0.30604                              \\
DCT@10\%                                                                                              & 0.11654                              & 0.17110                              & 0.19923                              & 0.26596                              \\

                                                                                                      &                                      &                                      &                                      &                                      \\
\textit{DWT@0.5\%}                                                                                    & \textbf{0.19874}                     & \textit{0.24449}                     & \textit{0.32192}                     & \textit{0.34581}                     \\
DWT@1\%                                                                          & \multicolumn{1}{l}{0.18257}          & \multicolumn{1}{l}{0.23025}          & \multicolumn{1}{l}{0.29205}          & \multicolumn{1}{l}{0.32320}          \\
DWT@5\%                                                                          & \multicolumn{1}{l}{0.15165}          & \multicolumn{1}{l}{0.20635}          & \multicolumn{1}{l}{0.26767}          & \multicolumn{1}{l}{0.24522}          \\
DWT@10\%                                                                         & \multicolumn{1}{l}{0.13050}          & \multicolumn{1}{l}{0.18475}          & \multicolumn{1}{l}{0.21481}          & \multicolumn{1}{l}{0.24519}          \\
\multicolumn{1}{l|}{}                                                                                 & \multicolumn{1}{l}{}                 & \multicolumn{1}{l}{}                 & \multicolumn{1}{l}{}                 & \multicolumn{1}{l}{}                 \\
\begin{tabular}[c]{@{}l@{}}intensity\\ +affine+elastic\end{tabular} & \multicolumn{1}{l}{0.17474} & \multicolumn{1}{l}{0.24239} & \multicolumn{1}{l}{0.31471} & \multicolumn{1}{l}{0.33533}\\
\textbf{\begin{tabular}[c]{@{}l@{}}DWT@0.5\\ +affine+elastic\end{tabular}} & \multicolumn{1}{l}{\textit{0.18721}} & \multicolumn{1}{l}{\textbf{0.26919}} & \multicolumn{1}{l}{\textbf{0.35432}} & \multicolumn{1}{l}{\textbf{0.38143}}
\end{tabular}
\caption{The F1-scores for the diseased class, averaged over 5 runs, are reported in this table. \emph{DCT@0.5\%}, for example, indicates that the maximum noise level, $\eta$ was 0.005 while synthesizing images using DCT, and is \textit{italicized} as it is the best performing spectral augmentation method. The results in \textbf{bold} indicate the best results obtained on this dataset.}
\label{resultstable}
\end{table}

The results depicted in Table \ref{resultstable} can be summarized as follows:

\textbf{Effect of training set size:} It can be ubiquitously observed that the U-net performs more favourably as the size of the training set increases.

\textbf{Baseline methods vs spectral transforms:} Intensity based augmentation does not fetch much benefit over simple copying, while elastic deformations and geometric transformations, perform favourably. The spectral transforms, namely DCT and DWT, which synthesize images by changing its texture, produce similar results to deformations and affine transformations, and suggest that the U-net benefits from learning to be invariant to texture changes in the lung tissue.

\textbf{Amount of noise in the spectral transforms:} We evaluate the performance of the spectral techniques at different levels of \begin{math} \eta \end{math}, the maximum corrupting noise. Results suggest that increasing the noise level causes a decrease in performance, as the signal to noise ratio increases.

\textbf{Combining different forms of augmentation:} We combine the best performing augmentation techniques by randomly applying affine transformations and elastic deformations to the images synthesized by \emph{DWT@0.5\%}. By combining affine, deformative and textural augmentations, the network can be taught to be invariant to all of them. This reflects in the performance of the network, as it outperforms one without any augmentation by 44.1\%, and the second-best by 10.3\%  F1-score.


\section{Conclusions}

In this work, we propose two novel data augmentation techniques based on frequency-domain transforms, and evaluate them on a lung-tissue segmentation task using a U-net architecture. Experiments suggest consistent favourable
performance of the proposed methods in relation to existing in-use data augmentation techniques. Furthermore, when combined with in-use methods, the best configuration of our method can significantly outperform every individual augmentation mechanism, and improve upon the simple replication F1-score for the diseased class by 44.1\%. We posit that these spectral methods create multiple synthetic views of data emulating multiple scans of the same subject being captured under varied CT-scan protocols, and can add complementary information in conjunction with existing data augmentation methods.

\section*{Acknowledgment}
This work was partially funded by the Netherlands Organisation for Scientific Research, Science Domain, VICI grant VI.C. 182.042.

\bibliographystyle{IEEEbib}
\bibliography{strings}

\end{document}